\documentclass[%
 reprint,%
 amssymb, amsmath,%
 aip,pop,%
]{revtex4-1}

\usepackage{bm}%
\usepackage[colorlinks=true,linkcolor=blue]{hyperref}%
\usepackage{graphicx}
\expandafter\ifx\csname package@font\endcsname\relax\else
 \expandafter\expandafter
 \expandafter\usepackage
 \expandafter\expandafter
 \expandafter{\csname package@font\endcsname}%
\fi
\begin{document}

\title{Velocity space degrees of freedom of plasma fluctuations}

\author{Sean Mattingly}%
\author{Fred Skiff}
\email{sean-mattingly@uiowa.edu}

\affiliation{212 Van Allen Hall, Department of Physics and Astronomy, University of Iowa, Iowa City, IA 52242}

\begin{abstract} 
	We present the first measurements of a plasma velocity-space cross-correlation
	matrix.  A singular value decomposition is applied to this inherently
	Hermitian matrix and the relation between the eigenmodes and the plasma
	kinetic fluctuation modes is explored. A generalized wave admittance is
	introduced for these eigenmodes.
\end{abstract}

\date{July 2017}%

\maketitle

Collective fluctuation modes of plasmas offer a general description of plasma
dynamics in collisionless \cite{Kampen1955,Case1959} and weakly
collisional plasmas\cite{NBS2004}. Kinetic modes have been studied both in low
density plasmas
\cite{Skiff2002,Souza-Machado1999,Diallo2005} and in fusion
plasmas, where in the latter they play an important role in the energetics of
electrostatic turbulence and
transport\cite{Hatch2011,Terry2006,Terry2014}.  Fluid and
magnetohydrodynamic descriptions capture only a few modes in the full plasma
collective mode spectrum. Understanding plasma fluctuations and transport
requires the inclusion of kinetic modes. However, kinetic modes are difficult
to isolate experimentally.

Detecting kinetic modes is best achieved by phase space resolving diagnostics. Here
we employ laser induced fluorescence (LIF)\cite{Stern1975} to
measure the plasma distribution fluctuation correlation function
\begin{equation}C(\vec x_1 , \vec x_2 , \vec v_1 , \vec v_2 ; \tau) = \langle
	\delta f(\vec x_1, \vec v_1, t) \delta f(\vec x_2, \vec v_2, t - \tau)
	\rangle_t, \label{eqn_C} \end{equation}
where $\langle \rangle_t$ denotes a time average and $\delta f = (f -
\langle f \rangle_t )$ is the phase space distribution function fluctuation. 

Earlier LIF measurements of $C$ found the autocorrelation given by the
diagonal $v_1 = v_2$.  Those measurements employed a single laser to
measure fluorescence at two separate points along the laser
beam\cite{Fasoli1994} and to measure $C$ as a function of single $v_1 = v_2$
with separation $x_1 - x_2$\cite{Diallo2005}.  Bicoherence spectra were also
derived from these measurements\cite{Uzun-Kaymak2006}. In this Letter, we
employ a local measurement technique with $x_1 = x_2$ and select two separate
and adjustable velocities $v_1$, $v_2$ so that a matrix of cross correlations
can be obtained. 

By doing this, we present in this Letter the first measurements of a plasma
velocity-space cross-correlation matrix. From this local measurement multiple
degrees of freedom can be isolated, including the kinetic modes. We validate
our noise reduction techniques through the symmetry properties of the
fluctuation correlation function. We demonstrate this technique on a weakly
coupled plasma and compute the associated eigenvectors in velocity space. A
frequency dependent generalized wave admittance can be derived for each
eigenmode. 

The locality of the measurement means that the technique may be applied to
plasmas in which a single-point velocity-sensitive measurement is possible and
multipoint measurements may be difficult.  Examples include \emph{in situ}
measurements of space plasmas, fusion plasmas, trapped plasmas
\cite{Anderegg1997}, and laser cooled plasmas \cite{Strickler}.

While LIF is frequently used to measure slowly varying moments of the ion
velocity distribution ($n$, $\vec{u}$, $T$, and higher), measuring
fluctuations as required in this experiment is difficult due to photon
statistics fluctuations that make a twofold contribution to noise. Firstly, the
LIF photon count rate is limited due to low metastable state densities and the
need to avoid excessive optical pumping. Secondly, a large fraction of the
light signal is not from single frequency LIF itself but rather from
electron collision-induced fluorescence.  Collision-induced fluorescence affects ions at
all velocities and is related to the metastable state populating mechanisms.
Thus, collision-induced fluorescence is linked to the signal magnitude. Therefore,
a statistical subtraction is needed.

Correlation measurements use ensemble averaging which permits evaluation and
elimination of photon statistics noise.  Nevertheless, there is only the above
mentioned handful of phase space incoherent fluctuation measurements.  For this
reason, the efficiency of subtraction of the background fluorescence is a
secondary question that we address in this Letter by exploiting symmetry
properties of this correlation function.

\begin{figure} \includegraphics{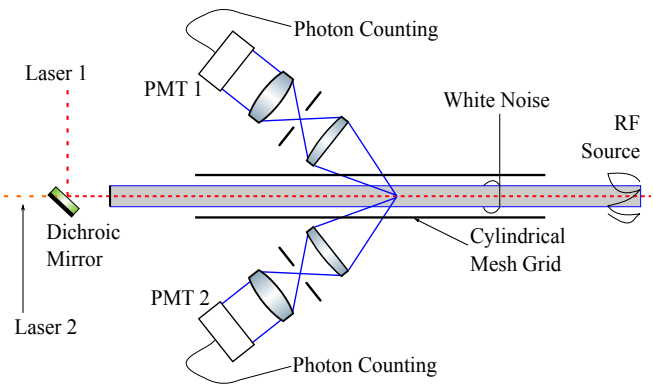} \caption{Experimental data
		acquisition scheme and setup. Two lasers, one at
		$668 \text{nm}$ and one at $611 \text{nm}$, are combined using a dichroic
		mirror and sent into the plasma. Corresponding to the two LIF
		schemes, PMT1 has a filter for 442nm while PMT2 has a filter
		for 461nm. An electrical white noise generator connected to a ring antenna
		excites the plasma modes with a broadband spectrum from
		$0 - 50$ kHz. A cylindrical mesh grid gives additional definition to radial
	mode structure.} \label{setup} 

\end{figure} 

The experiment is performed on an Argon II $13.56$ MHz RF inductively coupled
magnetized ($B \approx 0.067$ T) cylindrical plasma column of length $2.3$ m
and radius $r \approx 2.5$ cm\cite{Diallo2005}.  A Langmuir probe measures the
ion and electron densities $n \approx 9\times10^9~\text{cm}^{-3}$ and $T_e
\approx 10~\text{eV}$. LIF reveals $T_i \approx 0.08~\text{eV}$, though there
are significant deviations from a Maxwellian distribution. Ion neutral
collisions have frequency $\nu \approx 500$ Hz. The ambient ion acoustic wave and
drift wave amplitudes are above the thermal fluctuation level but are small due
to convective stabilization. Additionally, a ring antenna driven with white
noise excites broadband plasma modes in the system.

The experimental setup is shown in Fig \ref{setup}. Two separate lasers are
combined using a dichroic mirror and sent into the plasma to induce
fluorescence on two different metastable lines. These lines must be isolated -
otherwise transitions between the excited states can affect the laser optical
correlation. We have tested for this direct atomic collisional-radiative
connection through extensive searches for laser modulation frequency mixing,
which was not detected. The underlying form of these isolated LIF excitation
schemes are the same: laser 1 excites the ArII metastable state
${}^4\text{F}_{7/2}$ with 668 nm to ${}^4\text{D}^\circ_{5/2}$ which decays to
${}^4\text{P}_{3/2}$, emitting light at 442 nm; laser 2 excites
${}^2\text{G}_{9/2}$ with 611 nm to ${}^2\text{F}^\circ_{7/2}$ which decays to
${}^2\text{D}_{5/2}$, emitting light at 461 nm.  The dominant broadening
mechanism of the laser absorption spectrum is Doppler broadening and each laser
absorption spectrum is broadened by the same plasma ion distribution function
to within the noise level of 0.1\%.  Velocity selection is available with each
tunable laser since the laser bandwidth is $< 1~\text{MHz}$. Finally, to
suppress the Zeeman pattern corresponding to the left circular polarization,
these lasers are exclusively right circularized by passing them through a
Glan-Taylor laser polarizer and a quarter wave plate. The low field of $0.067$
T limits the width of each $\sigma^+$ polarized Zeeman group.

In this experiment, the collection optics are focused at the same point
with volume $0.20~\text{cm}^3$. Therefore, since the measurement scheme
spatially combines the lasers and then obtains the correlation function at this
physical volume, we obtain the correlation function of ions at two points
separated in velocity phase space $C(v_1, v_2, \tau)$.  

With this setup, first a full absorption spectrum is observed with each laser.
We then choose a set of velocities across the distribution function: the peak
of the distribution; $2/3$ the peak of the distribution; $1/2$ the peak; $1/3$
the peak; and one point on the tail. The ion velocity distribution function's
tail is due to how the plasma is produced \cite{Sarfaty1998}$^,$\cite{Skiff2000}.
In total, the same 8 velocity points are measured for each laser.

We then measure time series data $f(v, t)$ and $f(v', t)$ for the full array of
selected velocities. After demodulation with respect to $100$ kHz laser
chopping, the mean is subtracted to provide the fluctuation $\delta f = f  -
\langle f \rangle_t$.  Cross correlating and averaging $\delta f(v,t)$ and $\delta
f(v',t - \tau)$ with respect to $t$ gives $C(v, v', \tau)$ for each of the selected
velocities. This gives an $8\times8\times(2N-1)$ matrix where the velocities
form the first two axes and the time shift $\tau$ is the last axis.

During this process, photon statistics noise and background light are
suppressed at several points: light filtering in the set up; background light
subtraction through LIF signal demodulation (this step removes background light
correlations as well); filtering via a Gaussian windowing function of the time
cross correlation; and the suppression of photon statistics noise through
averaging. In order to validate this noise reduction, we examine the
correlation matrix.

Ideally, the matrix of correlation functions obeys the symmetry $C(v, v',
\tau)$ = $C(v', v, -\tau)$. However, this symmetry will not apply perfectly.
There are small errors in wavelength selection, and the applied magnetic field
induces slightly different Zeeman broadening for each laser absorption profile.
This breaks the velocity space symmetry.  However, keeping the magnetic
field at $0.067$ T ensures that the Zeeman subgroup is small compared to the
measurement spacing.  This is shown by the near symmetry of the actual raw data
matrix.

\begin{figure} \includegraphics{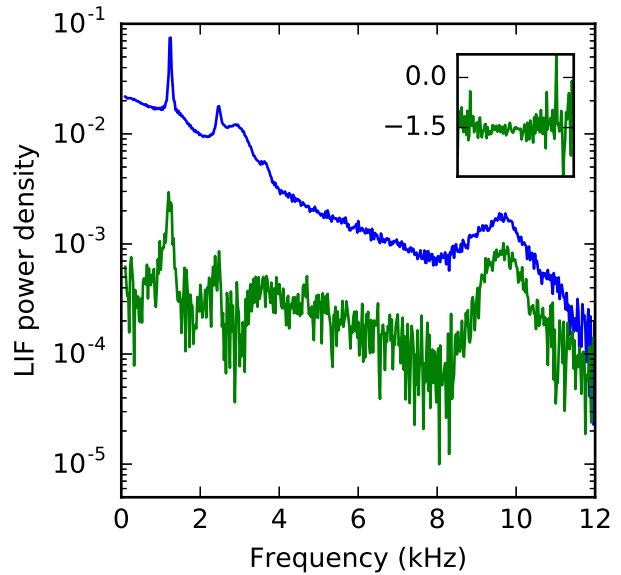}
        \caption{Representative power spectrum of a single velocity pair's time
                cross correlation.  The larger blue spectrum is the Hermitian
                component while the smaller green spectrum is the antihermitian
        component. The inset shows the phase of the antihermitian component
at the drift frequency and has the same abscissa as the outer figure.
} \label{herm_anti_spec} \end{figure}
Drift waves also help evaluate the degree of broken symmetry of the matrix.
Consider the physical set up: the periscopes shown in Fig.\ref{setup} are
oriented at $90^\circ$ to each other and each has two cones of light collection
volume with concurrent tips at the focus - where the lasers are located. The
drift wave amplitude peaks in the gradient region of the plasma - higher up in
the cone and away from the focus. With the physical setup in mind, we quantify
how the matrix symmetry can be broken. The strongest drift wave mode
corresponds to the first Fourier mode decomposition $e^{i m \phi}$ where
$m=1$\cite{Horton1999}. We check this by separating the matrix into symmetric
and antisymmetric components $S, A = \frac{1}{2}(C(v_i,v_j; \tau) \pm C(v_j,
v_i;-\tau))$ and taking the Fourier transform along the time axis to acquire
$\widetilde{S}, \widetilde{A}$.  These matrices, after this Fourier transform,
are Hermitian and antihermitian by construction.  Figure \ref{herm_anti_spec}
shows a representative power spectrum from the matrix and its unwrapped phase.
The unwrapped phase is $\pi / 2$, and since the periscopes are positioned to
have their light collection cones at angle $\pi / 2 $, this result is expected.
The drift wave represents a worst case of broken data matrix Hermiticity, and
the effect is mostly constrained in frequency space to the region around $f^*
\approx 10$kHz as shown in Fig.  \ref{herm_anti_spec}.

Figure \ref{herm_anti_spec}'s spectra show that the Hermiticity of the matrix
remains good to at least 10 dB for nearly all frequencies lower than the drift
wave. This validates our noise reduction processes for this
frequency range. This also provides a second validation that the Zeeman
broadening is not too large.

\begin{figure} \includegraphics{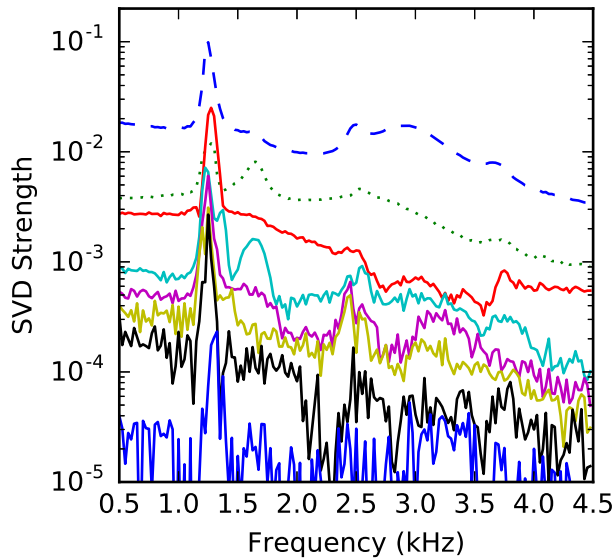}
        \caption{Power spectra from connecting the singular values of the
                Hermitian matrix.  A separate SVD is run on each 2D velocity
                space and connecting the singular values across frequency gives
                this plot. The fact that these power spectra differ is evidence
                of distinct modes. The third strongest mode, in solid red,
                becomes second strongest around $1250\text{Hz}$ and then drops
                back to being third strongest. These are the velocity space
        degrees of freedom of plasma fluctuations as a function of frequency.
} \label{singular_values}

\end{figure}
Obtaining the plasma degrees of freedom and the plasma modes is now possible.
We apply a singular value decomposition
(SVD)\cite{Nardone1992}$^,$\cite{Press2007} to the $2D$ Hermitian velocity
matrix $\widetilde{S}$ for every point in the frequency spectrum. The SVD gives
the effective rank, or detected degrees of freedom, of the matrix at a given
frequency by the number of singular values above the noise level. The magnitude
of each singular value determines the relative importance of its corresponding
principal axis. Assuming continuity of the principal axes in frequency, we
connect the singular values across the spectrum. The results are shown in Fig.
\ref{singular_values}.  The strongest singular value mode spectrum is
qualitatively similar to the undecomposed spectrum of Fig.
\ref{herm_anti_spec}. 

However, SVD does not come without drawbacks. It assumes
linearity and imposes the ansatz that the basis vectors are orthogonal. The
basis of the ion velocity space distribution function may not always fulfill
these assumptions. This is inextricably tied to SVD's strength: it does not
make any other assumptions about the form of these basis vectors - which is why
we use it here.

For comparison with the experiment, consider electrostatic ion waves where the
plasma is quasineutral and the electron density follows a Boltzmann
distribution. Rewriting the linearized Vlasov equation with a Poisson
bracket, expanding in the gyrophase coordinate $\varphi$ around the ion guiding
centers, using a BGK\cite{Bhatnagar1954} collision operator, and
integrating over the perpendicular velocity gives
\begin{eqnarray} 
	f_1(v_{||}) = \sum_{n=-\infty}^{\infty} e^{-k_\perp^2 v_{T}^2 /
\Omega^2} I_n(\frac{k_{\perp}^2 v_T^2}{\Omega^2}) \frac{n_1}{n_0} kT_e
\nonumber \\ \times \frac{i k_{||}\frac{\partial f_{0}}{\partial p_{||}} +
(\frac{i n \Omega}{kT_\perp} + \frac{\nu} {kT_e})f_{0}}{\nu + i k_{||} v_{||} -
i \omega - i n \Omega}, \label{eqn:modes} \end{eqnarray}
where $\Omega$ is the ion cyclotron frequency, $I_n$ is the modified Bessel
function of the first kind, and $\nu$ is the collision frequency.  At a given
frequency, then, there is a range of modes present in the plasma, each with its
own $k_{\perp}$ and $k_{||}$ corresponding to an $f_1$ given by
Eqn.~\ref{eqn:modes}. Similarly, in Fig.~\ref{singular_values}, SVD resolves a
subset of different modes for each given frequency and so we have separated
the different spatial plasma modes with a localized measurement.

We give two examples of spatial mode separation, the first without
Eqn.~\ref{eqn:modes} and the second with Eqn.~\ref{eqn:modes}. The chamber ion
acoustic longitudinal bounded mode is the $\approx 1250$ Hz peak and is strong
across all modes. A similar ion acoustic longitudinal mode bounded by the wire
mesh grid is at $\approx 1650$ Hz and is separated by SVD into the second and
fourth most important principal axes.

\begin{figure} 
        \includegraphics{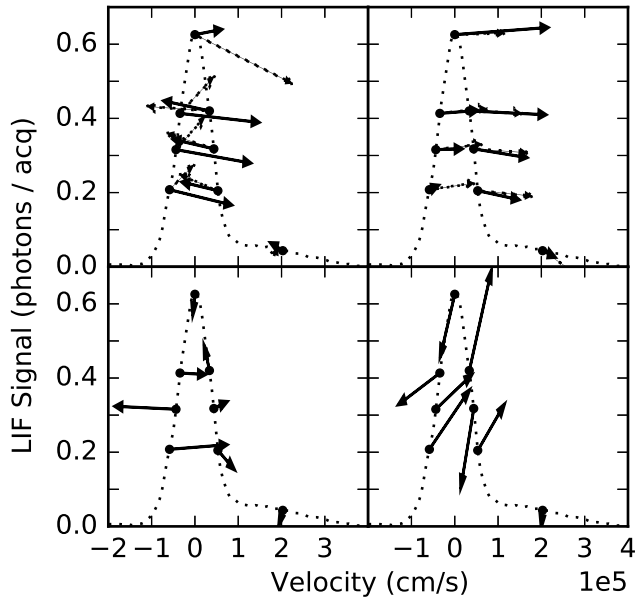}
        \caption{Eigenvectors of the four strongest singular value modes
                (strongest at top left and going clockwise) at $f = 800$ Hz.
                Each vector component is complex valued and represented with
                a solid Argand diagram with origin at the measurement point on
                the deconvolved ion velocity distribution function. The Argand
                diagrams from Eqn.~\ref{eqn:modes} are shown with dashed lines for the
                two strongest modes for comparison. The strongest
        mode's Argand diagram describes a traditional linearized ion acoustic
resonance.} \label{fig:single_eig} \end{figure}

The principal axes of the SVD provide a way to categorize these modes via
comparison with Eqn.~\ref{eqn:modes}. Since the matrix $\widetilde{S}$ is
Hermitian, SVD reduces to an eigenvector decomposition and thus the principal
axes are the complex valued eigenvectors of the ion velocity fluctuation
distribution function.  A set of these eigenvectors is displayed in
Fig.~\ref{fig:single_eig} for $f = 800~Hz$. By fitting these against
Eqn.~\ref{eqn:modes} evaluated with the measured plasma parameters, we can
categorize these modes. The resulting vectors from this fit are overlaid with
dotted Argand diagrams for the two strongest modes in
Fig.~\ref{fig:single_eig}. The second strongest mode at top right in
Fig.~\ref{fig:single_eig} corresponds to the spatially largest mode of the
plasma chamber with $\lambda_{||} = 460$ cm and fits best.  However, the
strongest mode itself corresponds to a smaller spatial mode with $\lambda_{||}
\approx 14.6$ cm and $\lambda_{\perp} \approx 1.4$ cm and does not fit as well.
This is another example of spatial mode separation from a local measurement and
that theory work is an avenue of future work to interpret this measurement
properly.

These mode structures do not remain constant as a function of frequency and may
also change in relative strength.  By minimizing the difference in eigenvectors
as frequency is varied, we can find the continuous path of evolution of
eigenvectors. Thus we determine if and when the relative strength of the mode
changes as a function of frequency. This process shows, for example, that the
second and third modes in Fig.~\ref{singular_values} switch in strength near
$1250$ Hz.

\begin{figure} \includegraphics{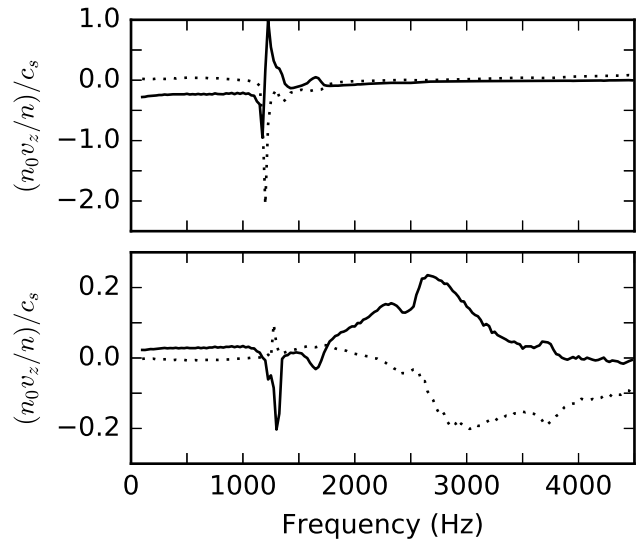}
	\caption{Normalized admittance of the eigenmodes as a function of
	frequency. Top is the strongest eigenmode while bottom is the second
strongest eigenmode. The real component is the solid line while the imaginary
is the dotted line.}
\label{fig:admit} 
\end{figure}

We introduce a generalized wave admittance by normalizing the appropriate
eigenmode.  In the case of the ion acoustic wave, combining the linearized
force balance equation with the ion acoustic wave dispersion relation gives
\begin{equation} \frac{n_0 v_z}{n} = c_s \sqrt{k_\perp^2 \frac{c_s^2}{\Omega^2
		- \widetilde{\omega^2}} + \frac{\omega}{\widetilde{\omega}}}
		\label{eqn:gen_admit} \end{equation}
where $\widetilde{\omega} = \omega - i \nu$.  The left hand side can be
calculated for each eigenmode by integrating the eigenvector to find the
denominator $n$ and the eigenvector's first velocity moment to find the numerator $n_0 v_z$.
Figure \ref{fig:admit} shows these admittances calculated using the two
strongest data eigenmodes normalized to the ion acoustic speed $c_s$.

The admittances of Fig.~\ref{fig:admit} show that the second mode, which corresponds
to modes constrained by the wire mesh grid, grows to reflect the larger $k_\perp$. The
peak in the first mode corresponds to the plasma chamber bounded mode.

To summarize, we have introduced a new method for measuring the degrees of
freedom and corresponding eigenmodes of plasma ion velocity distribution
function fluctuations. This method uses a singular value decomposition in order
to acquire the rank and the eigenmodes, in the case of a square Hermitian
matrix, or the principal axes more generally.  We show that this particular
localized measurement gives the discrete set of modes, both bounded modes and
kinetic modes, that constitute the plasma fluctuations. Analysis shows how
these modes change in strength as a function of frequency. We also calculate
a generalized wave admittance for each one of these eigenmodes.

The velocity space correlation function of the plasma distribution function
given by this method makes it possible to measure phase space fluctuation
spectra in terms of canonical velocity coordinates.  Refinement of this method
will be necessary since an integral transform to this end requires a nontrivial
kernel \cite{Morrison1994}.  A suggestive example is recent theoretical work
finding phase space density fluctuation spectra and electric field spectra
\cite{Morrison2008}.  Alternatively, it should be possible to project the data
matrix onto the basis of kinetic eigenmodes, which form a complete discrete
spectrum for fluctuations in a weakly collisional plasma \cite{NBS2004}.

This work is supported by the US DOE under the NSF-DOE program with grant number
DE-FG02-99ER54543.

%


\end{document}